\documentclass[12pt]{article}
\usepackage[margin=1in]{geometry}
\usepackage{mathrsfs}
\usepackage{amsmath}
\usepackage[T1]{fontenc}
\usepackage{subcaption}
\usepackage[font={footnotesize,it}]{caption}
\usepackage{authblk}
\usepackage{lineno}
\usepackage{graphicx}
\usepackage{fancyhdr}
\usepackage{hyperref}
\hypersetup{
  colorlinks=false,
  citecolor=blue
}

\usepackage[textfont=it, labelfont=bf]{caption}

\title{A mathematical model of national-level food system sustainability}
\author[1]{Conor Goold}
\author[2]{Simone Pfuderer}
\author[3]{William H. M. James}
\author[3]{Nik Lomax}
\author[4]{Fiona Smith}
\author[1]{Lisa M. Collins}
\affil[1]{\small{Faculty of Biological Sciences, University of Leeds, LS2 9JT, UK}}
\affil[2]{\small{School of Agriculture, Policy and Development, University of Reading, Reading, RG6 6AR, UK}}
\affil[3]{\small{School of Geography and Leeds Institute for Data Analytics, University of Leeds, LS2 9JT, UK}}
\affil[4]{\small{School of Law, University of Leeds, LS2 9JT, UK}}
\date{}

\begin{document}

\maketitle
\begin{abstract}
  The global food system faces various endogeneous and exogeneous, biotic and abiotic risk factors, including a rising human population, higher population densities, price volatility and climate change. Quantitative models play an important role in understanding food systems' expected responses to shocks and stresses. Here, we present a stylised mathematical model of a national-level food system that incorporates domestic supply of a food commodity, international trade, consumer demand, and food commodity price. We derive a critical compound parameter signalling when domestic supply will become unsustainable and the food system entirely dependent on imports, which results in higher commodity prices, lower consumer demand and lower inventory levels. Using Bayesian estimation, we apply the dynamic food systems model to infer the sustainability of the UK pork industry. We find that the UK pork industry is currently sustainable but because the industry is dependent on imports to meet demand, a decrease in self-sufficiency below 50\% (current levels are 60-65\%) would lead it close to the critical boundary signalling its collapse. Our model provides a theoretical foundation for future work to determine more complex causal drivers of food system vulnerability.\\
\end{abstract}


\section{Introduction}
Food security is defined as ``when all people at all times have physical, social and economic access to sufficient, safe and nutritous food to meet their dietary needs and preferences for an active and healthy life'' \cite{FAO2009}. The realisation of food security depends on the three pillars of access, utilisation and availability \cite{maxwell1996,barrett2010}, and therefore is an outcome of coupled agricultural, ecological and sociological systems \cite{hammond2012,ericksen2008,ingram2011}. In recent years, the resilience of food systems has become a priority area of research \cite{nystrom2019,tendall2015,bene2016,seekell2017} as biotic and abiotic, endogeneous and exogeneous demands on food systems grow, and the deleterious effects food systems currently have on the environment become more apparent \cite{springmann2018,strzepek2010}. The challenge of meeting the requirements of food security is for food systems to expand their production capacities while remaining resilient to unpredictable perturbations and limiting their effects on the environment, such as reducing waste \cite{ericksen2010}.

Food systems research is inherently transdisciplinary \cite{drimie2013,hammond2012}, of which one strand is computational and mathematical modelling. One utility of quantitative modelling is the ability to build and perturb realistic `systems models' of food systems to project important outcomes, such as future food production levels, farmer profitability, envirnomental degradation, food waste, and consumer behaviour \cite{springmann2018,marchand2016,sampedro2020,suweis2015,scalco2019,allen2016}. The difficulty in modelling food systems is their complexity, frequently resulting in large models with tens to hundreds of parameters and variables (e.g. \cite{sampedro2020,springmann2018}), which are challenging to analyse and even more challenging to statistically estimate from noisy real-world data \cite{sterman2000}. In contrast, a handful of authors have used relatively simple, theoretical models that are more amenable to formal analysis, and have fewer parameters to estimate from data. For example, \cite{suweis2015} link population dynamics to food availability and international trade using a generalised logistic model. \cite{tu2019} recently reported that the global food system is approaching a critical point signalling collapse into an unsustainable regime by condensing the multi-dimensional global food trade network into a bi-stable, one-dimensional system (using the framework of \cite{gao2016}). Simple models of coupled ecological, economic and agricultural processes have also been investigated, such as to explain the emergence of poverty traps, which have direct effects on individual's access to food \cite{ngonghala2017}.

While simplified models are less suitable for making predictions of complex systems' outcomes, their tractability makes them better placed to elicit causal explanations and generate hypotheses of how systems work \cite{smaldino2017,smaldino2019,otto2020}. These \textit{stylised} models are the backbones of scientific disciplines such as ecology \cite{may1973}, evolutionary biology \cite{boyd2003}, epidemiology \cite{kermack1927}, economics \cite{nerlove1958}, and physics \cite{strogatz1994}. For instance, stylised mathematical models of brain networks, animal collectives, ecosystems and cellular dynamics have been used to find the critical points at which they show abrupt qualitative changes in their behaviours \cite{sole1996,scheffer2001}. Food systems research, however, lacks such foundational models. Research on commodity production cycles has developed models to couple agricultural production, supply chains, consumer demand, price and human decision-making \cite{meadows1971,sterman2000}. However, applications of systems dynamics models of commodity cycles are frequently high dimensional, encoding multiple modes of behaviour that are not easily amenable to standard mathematical analysis. Gaining greater theoretical insight into the dynamics of food systems and food security would be aided by simpler models of coupled agri-food systems.

In this paper, we develop and analyse a stylised model of a food system inspired by the systems dynamics modelling of \cite{meadows1971} and \cite{sterman2000}, yet simple enough to offer general theoretical results. We focus on modelling a national-level food system, where the effects of international trade on domestic production is examined. Like the stylised models used to understand the causal processes in evolution, epidemics, and ecological interactions, our approach necessarily ignores many important features of real food systems. Nonetheless, its relative simplicity allows us to elucidate the precise conditions under which different stable modes of behaviour important to food system resilience emerge in our system.

We apply our theoretical model to the case of the UK pork industry, a key contributor to the UK meat industry which currently employs 75,000 employees and is worth \pounds1.25 billion \cite{DEFRA2019auk18}. Historically, pig industries have been of much interest to economists and agronomists as one of the first investigations into business or `pork' cycles \cite{haldane1934,coase1935,ezekiel1938,harlow1960,meadows1971,zawadzka2010,parker2014,sterman2000}. Business cycles reflect the oscillations between commodity prices and supply, which have been posited to be the result of both endogeneous (e.g. \cite{nerlove1958}) and exogeneous mechanisms (e.g. see \cite{gouel2012}). Over the last 20 years, however, the size of the UK pig industry has decreased by approximately 50\%, from 800,000 to approximately 400,000 sows, due to a combination of legislative, epidemiological, and trade-related issues \cite{taylor2006,dawson2009}. Following the ban on sow gestation crates in 1999, as well as disease outbreaks in the early 2000s, imports of pig meat increased by 50\% \cite{DEFRApigcattlestats2020}, exceeding domestic production. While domestic production has returned to accounting for around 60-65\% of total supply \cite{DEFRApigcattlestats2020}, the UK pig industry is still at risk from high costs of production \cite{BPEXprofitability2011} and `opportunistic dealing' within the pork supply chain favouring cheaper imports \cite{bowman2013}. The sustainability of the UK pig industry is, thus, a concern for UK food system resilience, particularly with the incipient threats of Brexit and the COVID-19 pandemic that are affecting international trade, labour availability, commodity prices, and consumer demand \cite{power2020,feng2017,poppy2019}. We demonstrate how our food systems model can be used to infer the sustainability of the UK pork industry using Bayesian estimation, enabling us to quantify full uncertainty in parameter estimtates.

\section{Materials and methods}

\subsection{Theoretical model}
\begin{figure}[t!]
  \centering
  \includegraphics[scale=1]{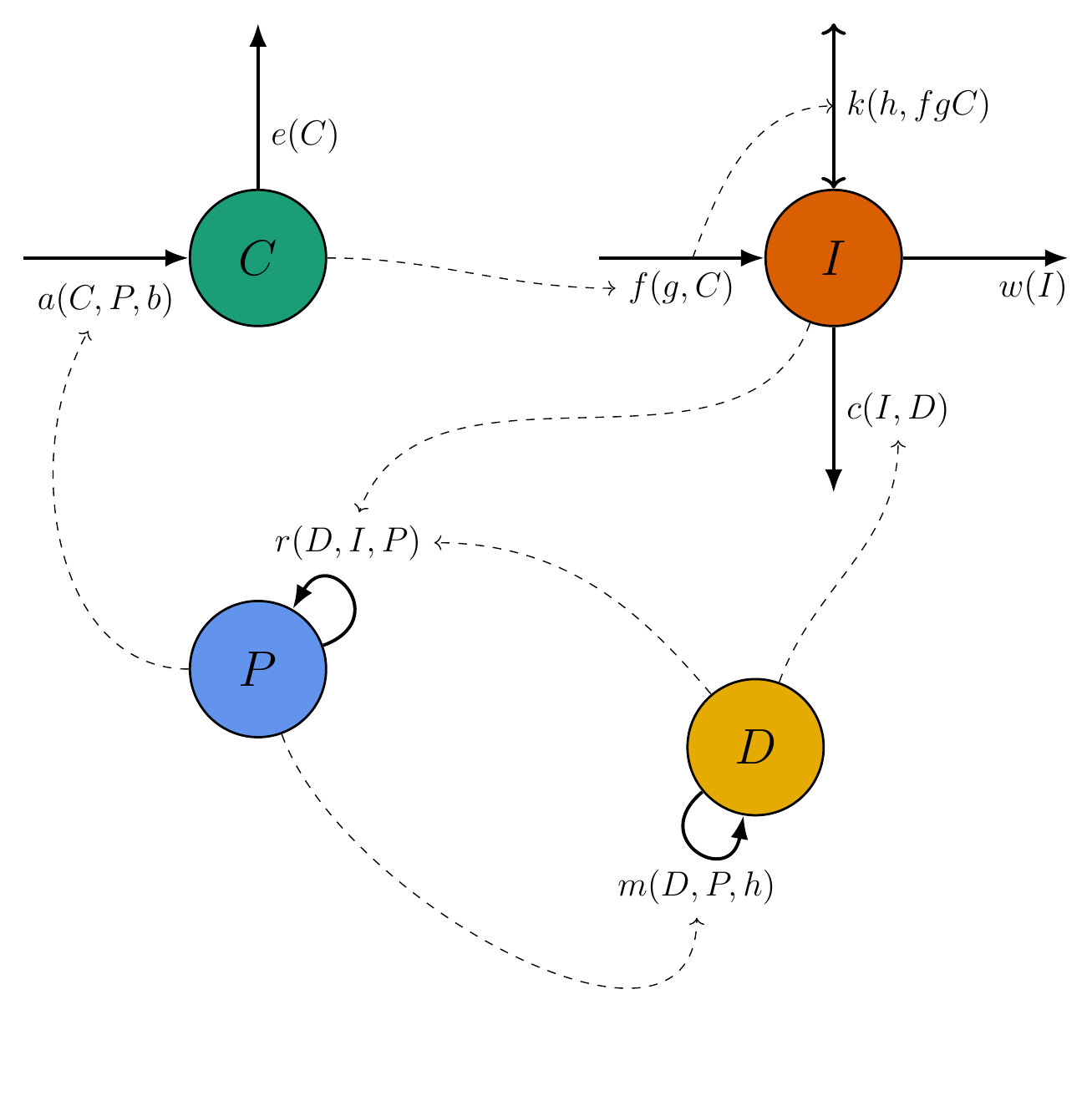}
  \caption{The general structure of the theoretical complex food system model. Blue circles denote the the four state variables (capital, inventory, demand and price). Solid arrows indicate the different flows into and out of each state variable comprising their rate of change, and the arrow labels display generic functions of the model state variables and parameters (see Table \ref{t_symbols} for the specific definitions used in this model). Dashed arrows show dependencies between different state variables and flows.}
  \label{fig_cfs}
\end{figure}

Our food system model is composed of coupled ordinary differential equations, with the state variables of capital, inventory, consumer demand, and price (Figure \ref{fig_cfs}; see variable and parameter definitions in Table \ref{t_symbols}). While we focus on food commodities, the model's generality means that it could be applied to other types of commodity. Capital represents a raw material used to gauge the viability of the domestic industry, which could represent, for instance, the number of animals in the breeding herd for meat industries (e.g. \cite{meadows1971}) or the number of paddy fields in rice supply chains (e.g. \cite{chung2018}). Inventory is the stock of processed food commodity being investigated. consumer demand represents the amount of inventory demanded per time unit by the population of consumers, and is dependent on the commodity price. The commodity price represents the price received by producers per unit of commodity produced, although we do not distinguish between producer and retail prices here (i.e. the producer price is assumed to be directly proportional to the retail price). While many of the mechanisms in supply chain functioning may be represented as a discrete-time system, we assume the aggregate behaviour of the national-level food system is adequately approximated in continuous time by a system of differential equations.

Capital changes according to the equation:
\begin{equation}
  \frac{dC}{dt} = a C \Big(\frac{P}{b} - 1\Big) - e C \quad, \quad C(0) = C_0
  \label{eq_capital}
\end{equation}
with initial condition at $t = 0$, $C_0$. The parameter $a$ is the rate of capital change (increase or decrease) depending on the price to capital production cost ($b$) ratio. Captial depreciates at rate $e$, where $e^{-1}$ is the average life-time of capital.

Inventory changes according to:
\begin{equation}
  \frac{dI}{dt} = f g C - w I - \frac{I}{sD + I} D + k (h - f g C) \quad, \quad I(0) = I_0
  \label{eq_inventory}
\end{equation}
The first term represents the amount of inventory generated by domestic capital per time unit (i.e. domestic supply), where $f$ is a production rate, and $g$ is a conversion factor representing the amount of inventory units produced per unit of capital. Inventory is wasted (i.e. produced but not consumed) at rate $w$. The third term denotes the rate of inventory consumption by consumers, which is a non-linear Holling type-II/Michaelis-Menten function asymptoting at $I$ for $D >> I$. The dimensionless function $I/(sD + I)$ can be interpreted as the proportion of the demand that can be satisfied with the current inventory level. The parameter $s$ is the `reference coverage' converting inventory demanded per time unit into commodity units, and is interpreted as the number of time units-worth of inventory processors desire to have in stock. Perishable food commodities (e.g. meat) will have a lower reference coverage, whereas less perishable items (e.g. rice, flour) can be stored for longer periods and, therefore, stock levels can be controlled by increasing $s$, lowering the proportion of demanded units satisfied.

The final term in equation \ref{eq_inventory} represents international trade, and its formulation can communicate different dynamics between domestic producers, processors and retailers. We retain simplicity by assuming that trade is proportional to the difference between a reference demand level ($h$) and current domestic production ($f g C$). When $h > f g C$, inventory is imported, and when $h < f g C$, inventory is exported. Realised demand ($D$) is a function of $h$ and the current commodity price (see below), and therefore $h$ represents the expected, baseline demand all else being equal \cite{sterman2000}. Trade levels adapt to the reference demand to avoid a positive feedback between higher prices, lower demand, and collapse of the commodity market for countries that are net importers, or a positive feedback between low prices, high demand, and exponentially increasing production for net exporters. In some industries, including the UK pork industry, cheaper international imports lower the domestic commodity price \cite{AHDBeuroexhange2015}, and thus importing more than domestic supply when demand drops due to higher prices is a mechanism for lowering the commodity price and increasing demand. The difference between reference demand and domestic production that is traded, however, is limited by factors such as trade tariffs (e.g. \cite{feng2017}) or the ability of a nation to attract trade partners, and thus the parameter $k$ controls the proportion of this difference. For countries that rely on international trade to supplement domestic supply to meet demand (net importers), $1 - k$ represents the self-sufficiency of the domestic industry (i.e. the percentage of total supplies produced domestically).

The instantaneous rate of change in demand is modelled as a simple function of reference demand and the commodity price to reference price ratio:
\begin{equation}
  \frac{dD}{dt} = m \Big( h \frac{q}{P} - D\Big) \quad, \quad D(0) = D_0
  \label{eq_demand}
\end{equation}
The parameter $m$ controls the time-responsiveness of demand. The reference price $q$ is typically interpreted as the price of substitute items \cite{sterman2000} or could represent consumers' overall willingness to pay. When the current price exceeds the reference price, demand falls, and vice versa.

Many models exist for describing the price of commodities (e.g. see \cite{legrand2019,deGoede2013} for some examples), and we adopt a relatively simple formulation here that has the rate of change of price depend only on the coverage:
\begin{equation}
  \frac{dP}{dt} = r P \Big(\frac{sD}{I} - 1\Big) \quad, \quad P(0) = P_0
  \label{eq_price}
\end{equation}
The coverage is a dimensionless quantity representing the amount of commodity needed to sustain current demand for $s$ time periods divided by the current inventory level. At rate given by $r$, the price increases when the coverage exceeds one, $sD/I > 1$, and decreases when coverage falls below $1$.

\begin{table}[t!]
  \centering
  \footnotesize
  \begin{tabular}{p{2cm}p{5cm}p{2cm}}
    \textbf{Symbol} & \textbf{Definition} & \textbf{Units} \\ \hline
    \textit{Variables} &&\\
    $C$    & Capital             & [$C$]\\
    $I$    & Inventory           & [$I$]\\
    $D$    & Demand              & [$It^{-1}$]\\
    $P$    & Price               & [$PI^{-1}$]\\
    $t$    & Time                & [$t$]\\
    \textit{Parameters} &&\\
    $a$    & Capital growth rate & [$t^{-1}$]\\
    $b$    & Cost of capital production & [$PI{-1}$]\\
    $e$    & Capital depreciation rate & [$t^{-1}$]\\
    $f$    & Capital production rate & [$t^{-1}$]\\
    $g$    & Capital conversion factor & [$IC^{-1}$]\\
    $w$    & Inventory waste rate     & [$t^{-1}$]\\
    $s$    & Reference coverage       & [$t$]\\
    $k$    & Trade strength        & [$-$]\\
    $h$    & Reference demand         & [$It^{-1}$]\\
    $m$    & Demand response rate     & [$t^{-1}$]\\
    $q$    & Reference price          & [$PI^{-1}$]\\
    $r$    & Price growth rate        & [$t^{-1}$]\\\hline
  \end{tabular}
  \caption{Symbols, definitions and their units for the complex food system model.}
  \label{t_symbols}
\end{table}

To make our model more generalisable, we non-dimensionalise the system of equations above (see supplmentary materials for non-dimensionalisation) using the dimensionless quantities in Table \ref{t_nd_symbols}, which reduces the number of parameters from 12 to 8. The non-dimensionalised system of equations is:

\begin{equation}
  \frac{dv}{d\tau} = v \Big(\alpha z - 1\Big) - \beta v
\end{equation}

\begin{equation}
  \frac{dx}{d\tau} = \delta v - \omega x - \gamma \frac{xy}{y + x} + \kappa ( \gamma - \delta v)
  \label{dimensionless_inventory}
\end{equation}

\begin{equation}
  \frac{dy}{d\tau} = \mu \Big( z^{-1} - y\Big)
\end{equation}

\begin{equation}
  \frac{dz}{d\tau} = \rho z\Big(\frac{y}{x} - 1\Big)
\end{equation}
where $\{v, x, y, z\}$ now represent the dimensionless state variables, $\tau$ is rescaled time, and the dimensionless parameter groups are denoted by Greek letters.

\begin{table}[b!]
  \centering
  \footnotesize
  \begin{tabular}{lll}
    \textbf{Symbol} & \textbf{Definition} & \textbf{Description} \\ \hline
    \textit{Variables} &\\
    $v$  & $\frac{C}{C_0}$    & Rescaled capital        \\
    $x$  & $\frac{I}{hs}$  & Rescaled inventory      \\
    $y$  & $\frac{D}{h}$  & Rescaled demand             \\
    $z$  & $\frac{P}{q}$  & Rescaled price               \\
    $\tau$  & $\frac{t}{1/a}$  & Rescaled time\\
    \textit{Parameters} &\\
    $\alpha$   & $q/b$  & Reference profitability\\
    $\beta$    & $e/a$  & Capital replacement-depreciation ratio\\
    $\delta$    & $f g C_{0}/(ahs)$ & Initial production-demand ratio\\
    $\omega$    & $w/a$ & Waste-production ratio\\
    $\gamma$    & $1/(as)$ & Capital replacement-coverage ratio\\
    $\kappa$    & $k$    & Trade strength \\
    $\mu$       & $m/a$  & Demand response-capital replacement ratio\\
    $\rho$      & $r/a$  & Price response-capital replacement ratio\\
    \hline
  \end{tabular}
  \caption{Symbols and definitions for the dimensionless complex food system model.}
  \label{t_nd_symbols}
\end{table}

\subsection{Data sources}
A range of data is collected on the UK pork industry, but raw time series data is only available for certain variables and time frames. To fit our theoretical model, we focused on monthly data over a period of 5 years from 2015 through 2019, which covers the available annual data for the `All pig price' per kilogram of deadweight (i.e. a combined price for standard and premium pigs). All data sources used to fit the model are presented in Table \ref{table_data_sources}. Monthly data for the inventory of pork, taking into account current levels of consumption and waste (e.g. the amount held in cold storage), is not reported in the UK. However, as an approximation, we used the total new monthly supplies, calculated as domestic production of pig meat plus imported pig meat minus exported pig meat. No data is available on consumer demand, as this is a theoretical quantity. Missing data was considered missing completely at random (i.e. ignorable) because data collection schemes are largely independent and fixed. For instance, missing breeding herd data were not considered dependent on the price or new supplies data.

\begin{table}[]
  \centering
  \footnotesize
  \begin{tabular}{p{1.5cm}p{7cm}p{6cm}}
    \textbf{Variable} & \textbf{Data} & \textbf{Details} \\ \hline
    $t$ & Time set to monthly intervals between 2015 through 2019 & Price data only available for this time period \\
    $C$
    & Number of female pigs in the breeding herd (June and December surveys) \cite{DEFRAlivestocknumbers}
    & The breeding herd represents the main capital of meat industries \\
    $I$
    & Amount (kg) of new pork available for consumption \cite{DEFRApigcattlestats2020,AHDBpigmeatrade}
    & Calculated as UK production (from \cite{DEFRApigcattlestats2020}) plus imports and minus exports (from \cite{AHDBpigmeatrade})  \\
    $D$ & No data available & Demand is a latent quantity \\
    $P$ & All pig price (kg/deadweight) \cite{DEFRAlivestockprices} & The price producers receive, assumed to be proportional to the retail price \\ \hline
  \end{tabular}
  \caption{UK pork industry data sources used to fit the food systems model}
  \label{table_data_sources}
\end{table}

\subsection{Bayesian estimation}
The parameters and initial conditions of the non-dimensionalised model were estimated using Bayesian estimation in the probabilistic programming language Stan \cite{carpenter2017} using the RStan interface in R \cite{stan2019,rcoreteam2020} using Stan's Runge-Kutta 4th and 5th order integration scheme (see Stan code in the supplementary materials). The available monthly time series data, $Y$, for month $i$ and state variable $j$ was assumed log-normal distributed (to ensure positivity):

\begin{equation}
  Y_{i}^{j} \sim \text{Lognormal}( ln( Z^{j} ), \sigma^{j})
\end{equation}
where $Z^{j}$ is the state variable computed from the food systems model. In addition to fitting the state variables of the model to the time series data, we fit the UK monthly production figures, and the monthly imports and exports, to the respective flows from the model:

\begin{equation}
  \text{Production} \sim \text{Lognormal}( ln( f g Z^{1} ), \epsilon_{1})
\end{equation}

\begin{equation}
  \text{Imports} \sim \text{Lognormal}( ln( k h ), \epsilon_{2})
\end{equation}

\begin{equation}
  \text{Exports} \sim \text{Lognormal}( ln( k f g Z^{1} ), \epsilon_{3})
\end{equation}
To aid computation, all parameters were transformed to a similar scale and given standard unit normal prior distributions (see full model specification in the supplementary materials), and were back-transformed to the appropropriate scale when integrating the model. We did not estimate the parameters $b$ (cost of capital production) and $g$ (conversion factor from capital to inventory units) because these were known with enough certainty beforehand: $b$ was set to the 138.3 p/kg (the average cost of production between 2015 and 2020), and $g$ was set to 82.4 kg/pig, reflecting the average slaughter weight of pigs (109.9kg) multiplied by a 75\% dressing yield (0.75) most recently reported by \cite{AHDBpocketbook2018}. We ran 4 Markov chain Monte Carlo (MCMC) chains consisting of 2,500 iterations of warmup and 2,500 iterations of sampling, providing 10,000 samples from the posterior distribution for inference. All chains ran without any divergent transitions, and all parameters had effective sample sizes $>> 1000$ and $\hat{R}$ statistics (i.e. the Gelman-Rubin diagnositc) $0.99 < \hat{R} < 1.002$ indicating convergence. Each parameter is summarised by its mean and 95\% highest density interval (HDI, the most 95\% most likely values). All data and code are available at \href{https://github.com/cmgoold/cfs-model}{https://github.com/cmgoold/cfs-model}.

\section{Results}

\subsection{Mathematical analysis}
No explicit solutions to the four-dimensional system of non-linear equations exist. Nonetheless, its dynamics can be summarised by investigating its stable modes of behaviour. To investigate stability, we conduct linear stability analyses \cite{strogatz1994}. Linear stability analysis is based on a Taylor series expansion in multiple variables around the fixed points ($\{\hat{v}, \hat{x}, \hat{y}, \hat{z}\}$), where asymptotic (i.e. $t \rightarrow \infty$) stability to small perturbations can be inferred when the real part of the eigenvalues of the system's matrix of partial derivatives (the Jacobian matrix representing the linearisation around the fixed points) evaluated at each equilibrium are negative. Notably, the inverse of the leading eigenvalue of the Jacobian matrix determines the `characteristic return time' of the system, with more resilient systems returning more quickly to their equilibria following a disturbance \cite{pimm1984}.

\subsubsection{Without international trade}
When international trade is not present (i.e. $\kappa = 0$), there is one stable fixed point of the system given by the state variable values $\big \{\frac{\alpha(2\omega + \gamma)}{2 \delta (1 + \beta)}, \frac{\alpha}{1 + \beta}, \frac{\alpha}{1 + \beta}, \frac{1 + \beta}{\alpha}\big\}$. Another fixed point is where all state values are 0 (i.e. no industry). Conducting a linear stability analysis around the latter fixed point, the eigenvalues of the Jacobian matrix at this equilibria are $\{(-1 - \beta), - \omega, - \mu, - \rho)\}$. All parameters are defined to be positive (except $\kappa$, which is $0$ here), meaning there are no conditions where the `no industry' equilibria will be stable when international trade is absent. In other words, the domestic industry is always viable.

\subsubsection{With international trade}
When $0 < \kappa < 1$, international trade is possible, opening the possibility of competition between domestic and international products, or of an export market for domestic production. An unstable equilibria still exists where all state variables are 0, except for $\hat{x} = \frac{\kappa \gamma}{\omega}$ because importing products is possible. However, in addition, there is 1) an unsustainable domestic production equilibrium, where the food system is reliant on international imports, and 2) a sustainable domestic production equilibrium, where the domestic industry co-exists with international trade.

\begin{figure}[t!]
  \begin{subfigure}{0.5\textwidth}
      \includegraphics[scale=0.6]{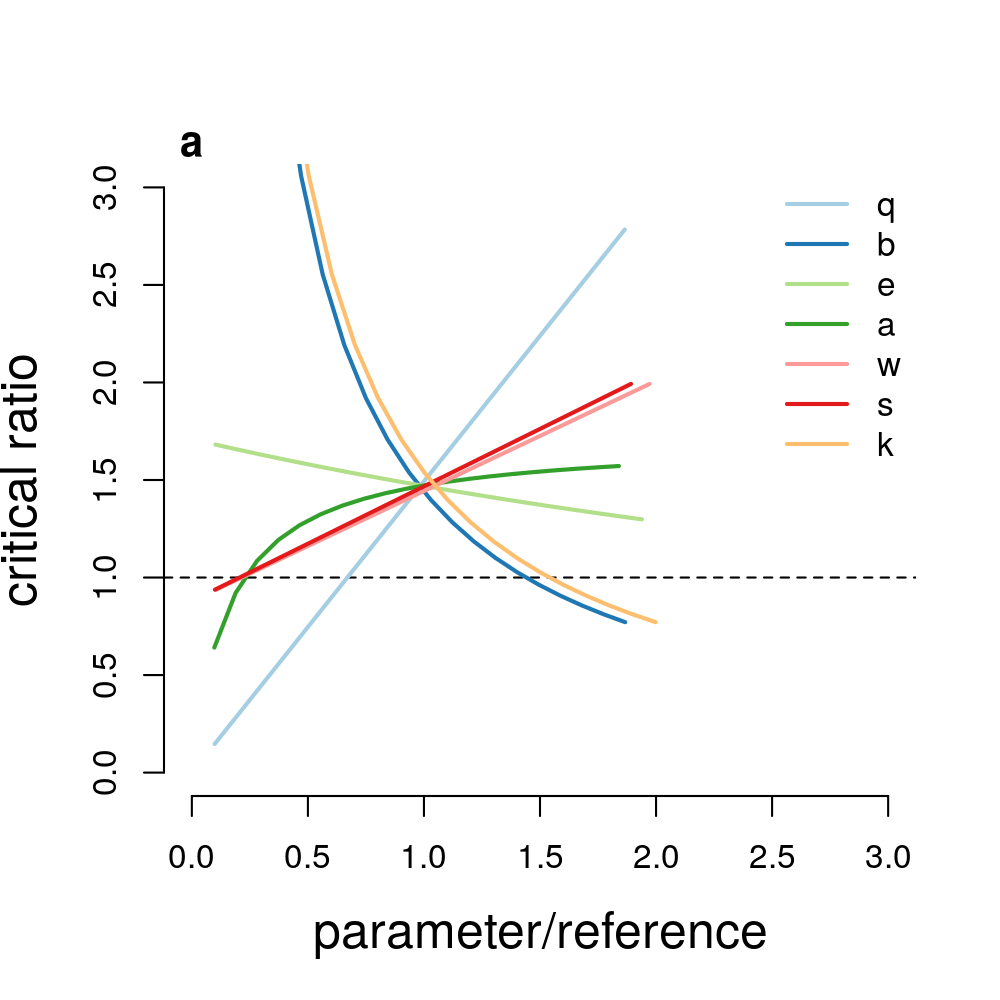}
  \end{subfigure}%
  ~%
  \begin{subfigure}{0.5\textwidth}
      \includegraphics[scale=0.6]{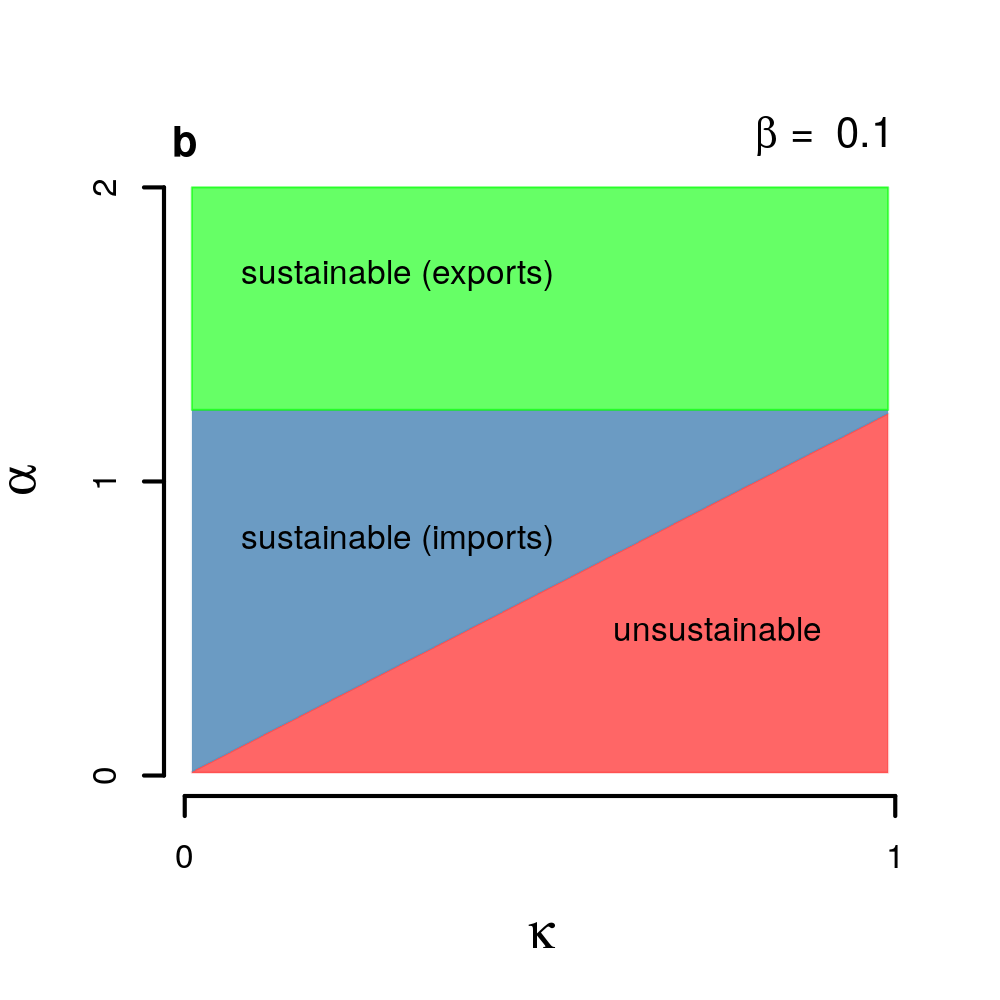}
  \end{subfigure}

  \begin{subfigure}{0.5\textwidth}
      \includegraphics[scale=0.6]{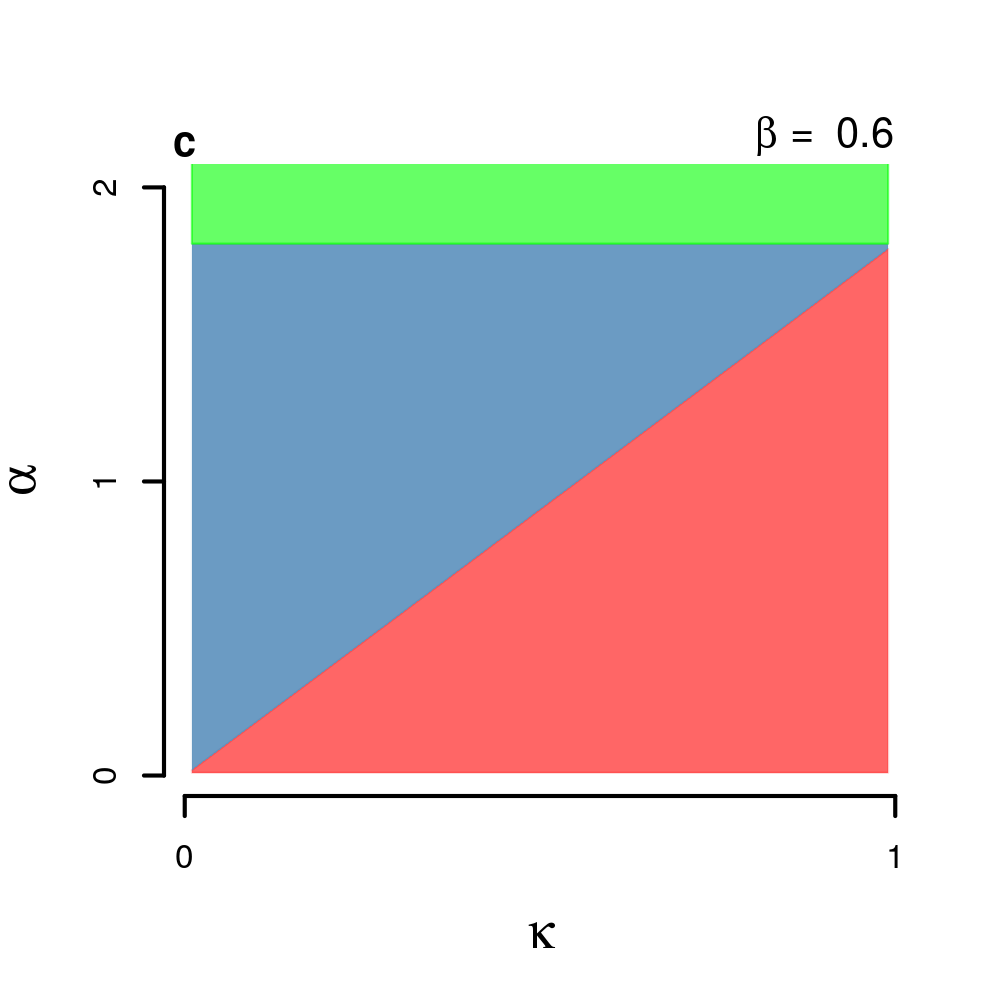}
  \end{subfigure}%
  ~%
  \begin{subfigure}{0.5\textwidth}
      \includegraphics[scale=0.6]{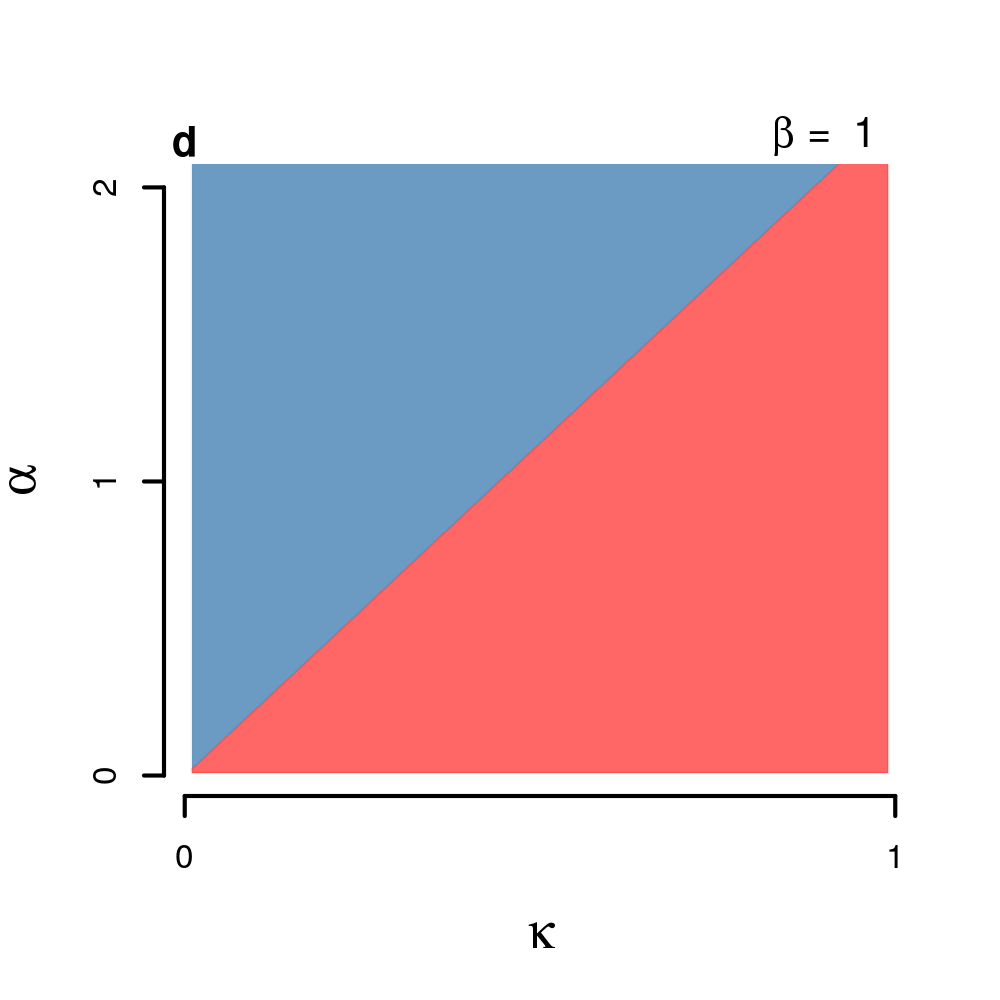}
  \end{subfigure}%

  \caption{Stability of the model incorporating international trade ($0 < \kappa < 1$). Panel a shows the sensitivity of the critical ratio to parameters in Table \ref{t_symbols}. The x-axis shows the ratio of the parameters to their reference values ($q = 160$, $b = 140$, $e = 0.033$, $a = 0.2$, $w = 0.33$, $s=1$, $k=0.5$). The horizontal dashed line shows the critical ratio threshold of unity. Panels b-d show the stable modes of behaviour in ($\kappa, \alpha$) space for differing values of $\beta$, distinguishing between unsustianable (red), sustainable with imports (blue), and sustainable with exports (green) behaviours. Panels b-d are produced with the remaining parameters at $\gamma = 26$, $\omega = 10$ and $\delta = 5$.}
  \label{figure2}
\end{figure}

The unsustainable domestic production equilibria is given by the set of fixed points $\{0, \frac{\kappa \gamma}{\omega + \frac{\gamma}{2}}, \frac{\kappa \gamma}{\omega + \frac{\gamma}{2}}, \frac{\omega + \frac{\gamma}{2}}{\kappa \gamma}\}$. The Jacbobian matrix ($\boldsymbol{J}$) at this equilibria evalutes to:

\begin{equation}
  \boldsymbol{J} \Big |_{\big \{0, \frac{\kappa \gamma}{\omega + \frac{\gamma}{2}}, \frac{\kappa \gamma}{\omega + \frac{\gamma}{2}}, \frac{\omega + \frac{\gamma}{2}}{\kappa \gamma}\big \}} =
  \begin{pmatrix}
    \frac{\alpha(\omega + \frac{\gamma}{2})}{\kappa \gamma} - 1 - \beta   &    0     &     0    &  0 \\
    \delta (1 - \kappa) & - \omega - \frac{\gamma}{4}  &  - \frac{\gamma}{4} & 0 \\
    0      &           0            & -\mu     & - \mu (\frac{\kappa \gamma}{\omega + \frac{\gamma}{2}})^2\\
    0      &   - \rho (\frac{\omega + \frac{\gamma}{2}}{\kappa \gamma})^2    & \rho (\frac{\omega + \frac{\gamma}{2}}{\kappa \gamma})^2 & 0 \\
  \end{pmatrix}
\end{equation}
and its eigenvalues ($\boldsymbol{\lambda}$) are the roots of the fourth-degree characteristic polynomial:

\begin{equation}
  \Big(\frac{\alpha(\omega + \frac{\gamma}{2})}{\kappa \gamma} - 1 - \beta - \lambda\Big) \Big[ - \lambda^3 + \big(-\omega - \frac{\gamma}{4} - \mu\big) \lambda^2 + \big(- \mu(\omega + \rho) - \frac{\gamma}{4}\big) \lambda - \frac{\gamma}{2} \mu \rho\Big] = 0
\end{equation}
The first eigenvalue can be determined directly as:

\begin{equation}
  \lambda_{1} = \frac{\alpha(\omega + \frac{\gamma}{2})}{\kappa \gamma} - 1 - \beta
\end{equation}
By using the Routh-Hurwitz conditions, the sign of the remaning eigenvalues' real parts \cite{ottoday2011} will always be negative (see the supplementary materials). Ultimately, the unsustainable domestic production mode will be stable if:

\begin{equation}
  \text{critical ratio} = \frac{ \alpha (\omega + \frac{\gamma}{2})}{\kappa \gamma(1+\beta)} < 1
\end{equation}
The dependence of this critical ratio to changes in the original parameter values is shown in Figure \ref{figure2}a. The numerator represents a weighting of three factors: $i)$ the \textit{reference profitability} of capital production ($\alpha$; Table \ref{t_nd_symbols}), $ii)$ the \textit{need for new commodity}, where either higher rates of waste or greater reference coverage increases the critical ratio, and $ii)$ the \textit{speed of capital production}, where higher rates ($a$ in Table \ref{t_symbols}) increases the critical ratio (Figure \ref{figure2}a). By contrast, the denominator represents the \textit{total strength} of international trade, which is composed of the trade strength parameter ($\kappa$), weighted by the viability of domestic capital: if the capital production rate increases, or capital depreciation rate decreases, the total trade strength becomes smaller, serving to increase the critical ratio.

When the critical ratio exceeds 1, the sustainable domestic production equilibrium is given by $\{\frac{2 \gamma \kappa (- 1 - \beta) + \alpha (\gamma + 2 \omega)}{2 \delta (1+\beta)(1 - \kappa) }, \frac{\alpha}{1 + \beta}, \frac{\alpha}{1 + \beta}, \frac{1 + \beta}{\alpha}\}$. In the latter case, the equilibrium values of inventory, demand and price are the same as the equilibrium values of the model without international trade (see above), determined by the profitability of the domestic industry and the ratio of capital depreciation and growth rates. However, equilibrium inventory and demand are lower, and equilibrium price is higher, than when domestic supply is unsustainable. For example, the conditions for equilibrium price when domestic supply is unsustainable to be higher than when domestic supply is sustainable (i.e. $\frac{\omega + \frac{\gamma}{2}}{\kappa \gamma} > \frac{1 + \beta}{\alpha}$) is exactly the critical ratio. The latter trends are also seen the closer a system comes to the critical ratio. Thus, while international trade in the short term increases inventory levels, decreases the coverage (see model description) and, therefore, decreases the price and increases demand, the long-term result of the unsustainable domestic production regime is higher prices, lower demand and lower inventory levels.

The equilibrium value of capital is similar to the equilibrium value found in the model without international trade, but now factors in trade strength, $\kappa$ (both equilibria are equal when $\kappa = 0$). Specifically, whether $\kappa$ has a positive or negative influence on the long-term sustainable equilibrium domestic capital depends on whether the system is characterised by net imports (domestic supply is less than reference demand) or net exports (domestic supply exceeds reference demand), i.e. whether $\frac{\gamma}{\delta} - \hat{v}$ in equation \ref{dimensionless_inventory} is greater than or less than zero. If domestic supply is less than reference demand, and the critical ratio exceeds 1, increasing trade strength will reduce the equilibrium value of capital in the long term limit (i.e. as $\tau \rightarrow \infty$). However, if domestic supply exceeds reference demand, increasing trade strength will increase the equilibrium capital due to a greater ability to export surplus product. From the previous inequality, we can define the surplus ratio, which signals that domestic supply will be greater than reference demand (net exports) if:

\begin{equation}
  \text{surplus ratio} = \frac{\alpha (\omega + \frac{\gamma}{2})}{\gamma (1 + \beta)} \equiv \kappa \cdot \text{critical ratio} > 1
\end{equation}
which equals the critical ratio cancelling out the trade strength. Figures \ref{figure2}b-d demonstrate the relationship between the sustainable and unsustainable stable modes of behaviour in $(\kappa, \alpha)$ space for differing values of $\beta$, as well as the distinction between the sustainable state characterised by net imports or exports.

\subsection{Application to UK pig industry}
The critical ratio for the UK industry was estimated to be credibly above 1 (Table \ref{table_parameter_estimates}), suggesting the industry is in a sustainable condition according to this model. The reference demand is estimated to be approximately 1.6 times that of the UK estimated pig meat annual consumption (approximately 140 million kg based off 25 kg/person/year and a population size of 66.65 million people; \cite{AHDBpocketbook2018}). The trade strength is approximately 0.36 on average, which is consistent with the current self-sufficiency level of around 65\% (i.e. around 35\% of UK pig meat is imported). The difference between $\alpha$ and the $\text{surplus ratio}$ is credibly less than zero (mean: -0.64; HDI: [-0.83, -0.43]), reflecting that UK production of pig meat does not meet expected demand. The critical $\kappa$ value needed to push the UK domestic industry into the unsustainable regime is 0.61 (95\% HDI: [0.56, 0.65]).

\begin{figure}[t!]
  \begin{subfigure}{0.5\textwidth}
    \includegraphics[scale=0.5]{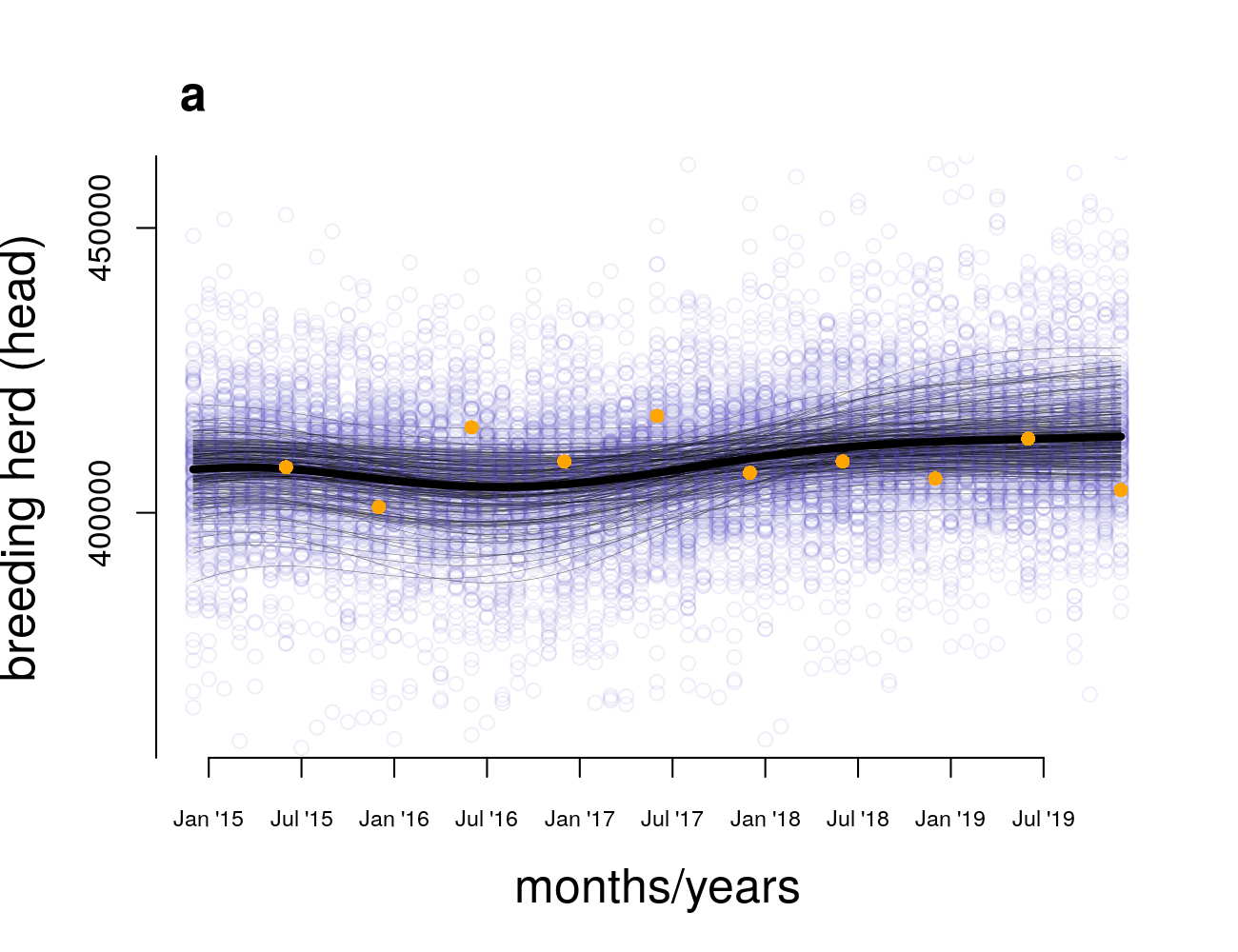}%
  \end{subfigure}%
  \begin{subfigure}{0.5\textwidth}
    \includegraphics[scale=0.5]{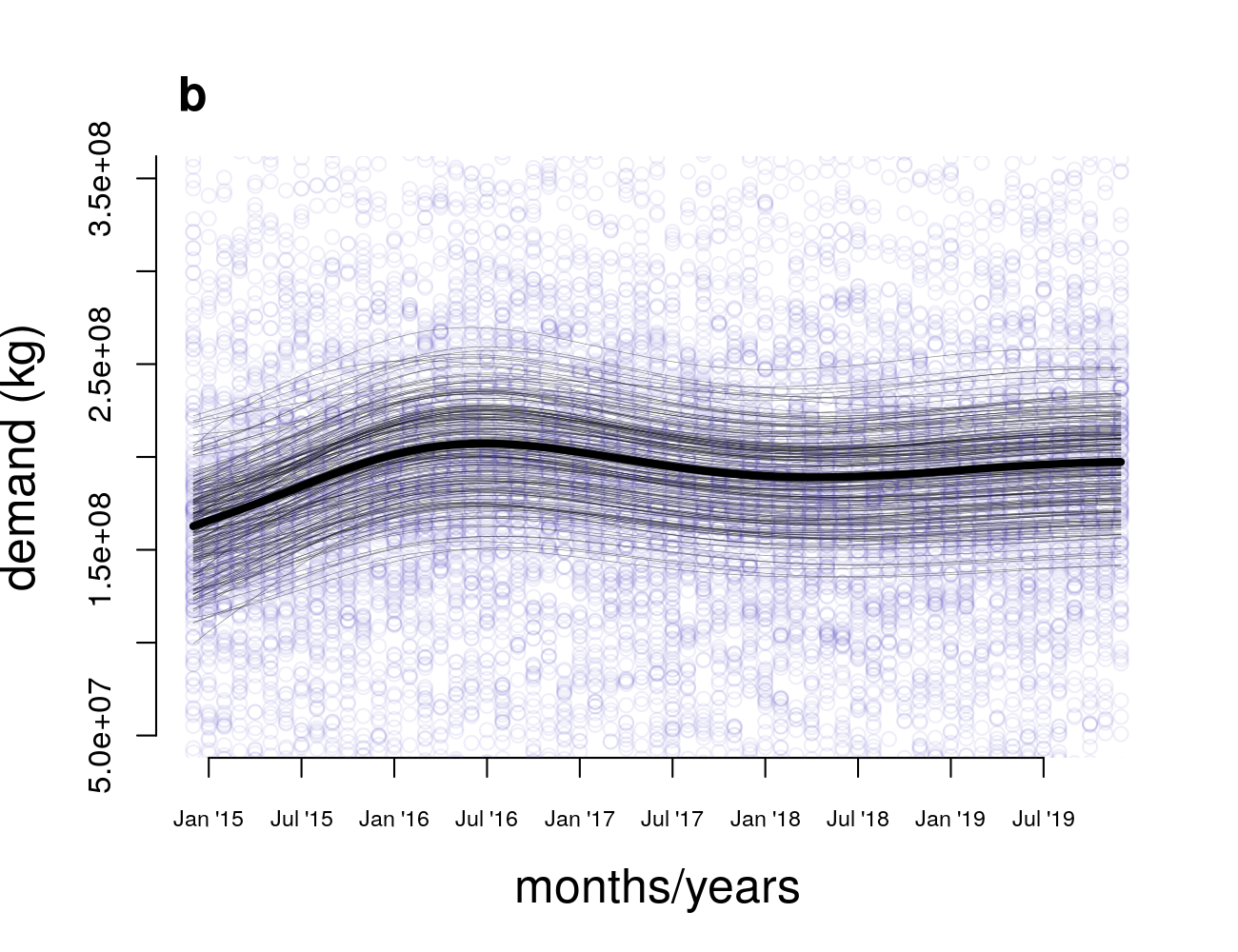}
  \end{subfigure}

  \begin{subfigure}{0.5\textwidth}
    \includegraphics[scale=0.5]{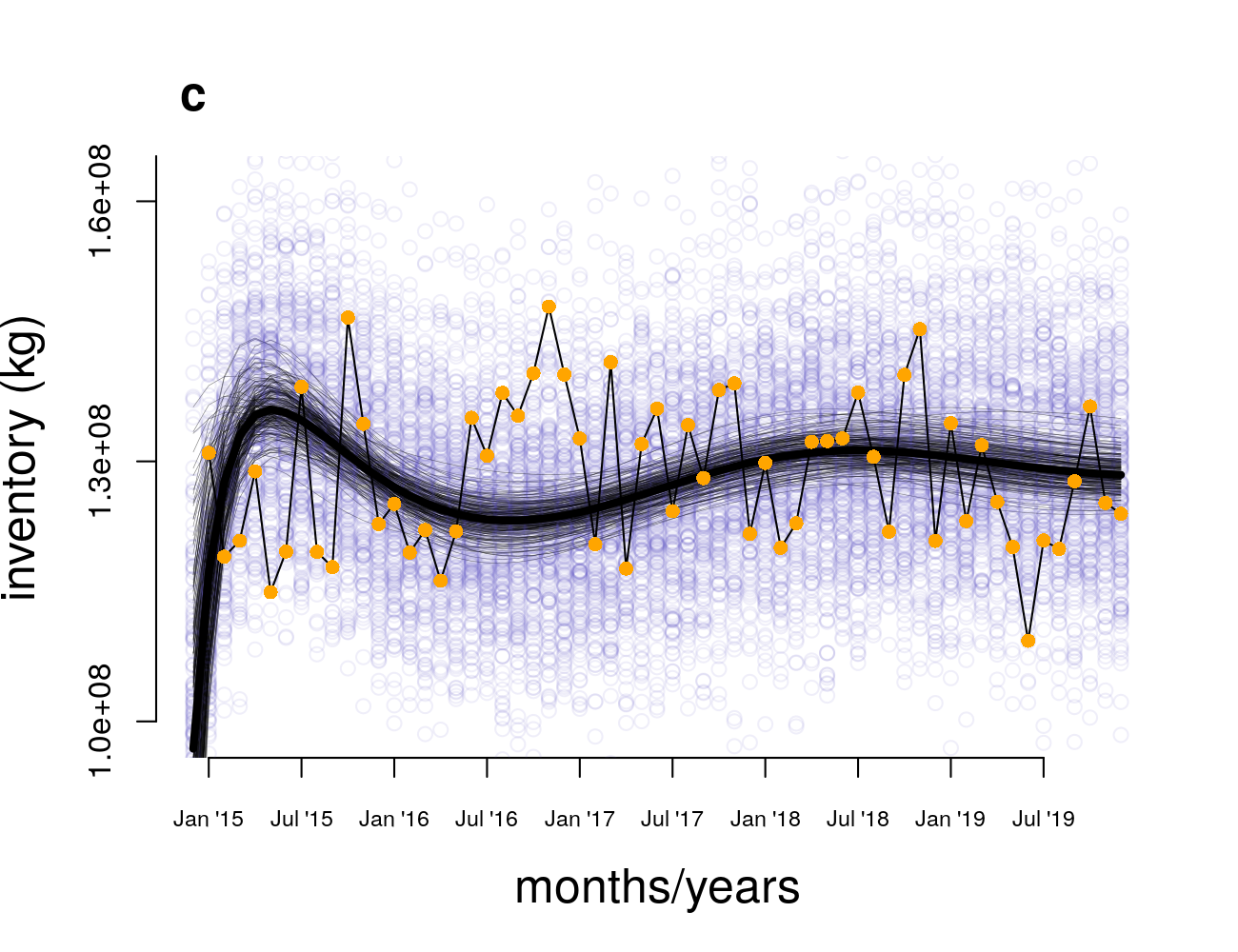}
  \end{subfigure}%
  \begin{subfigure}{0.5\textwidth}
    \includegraphics[scale=0.5]{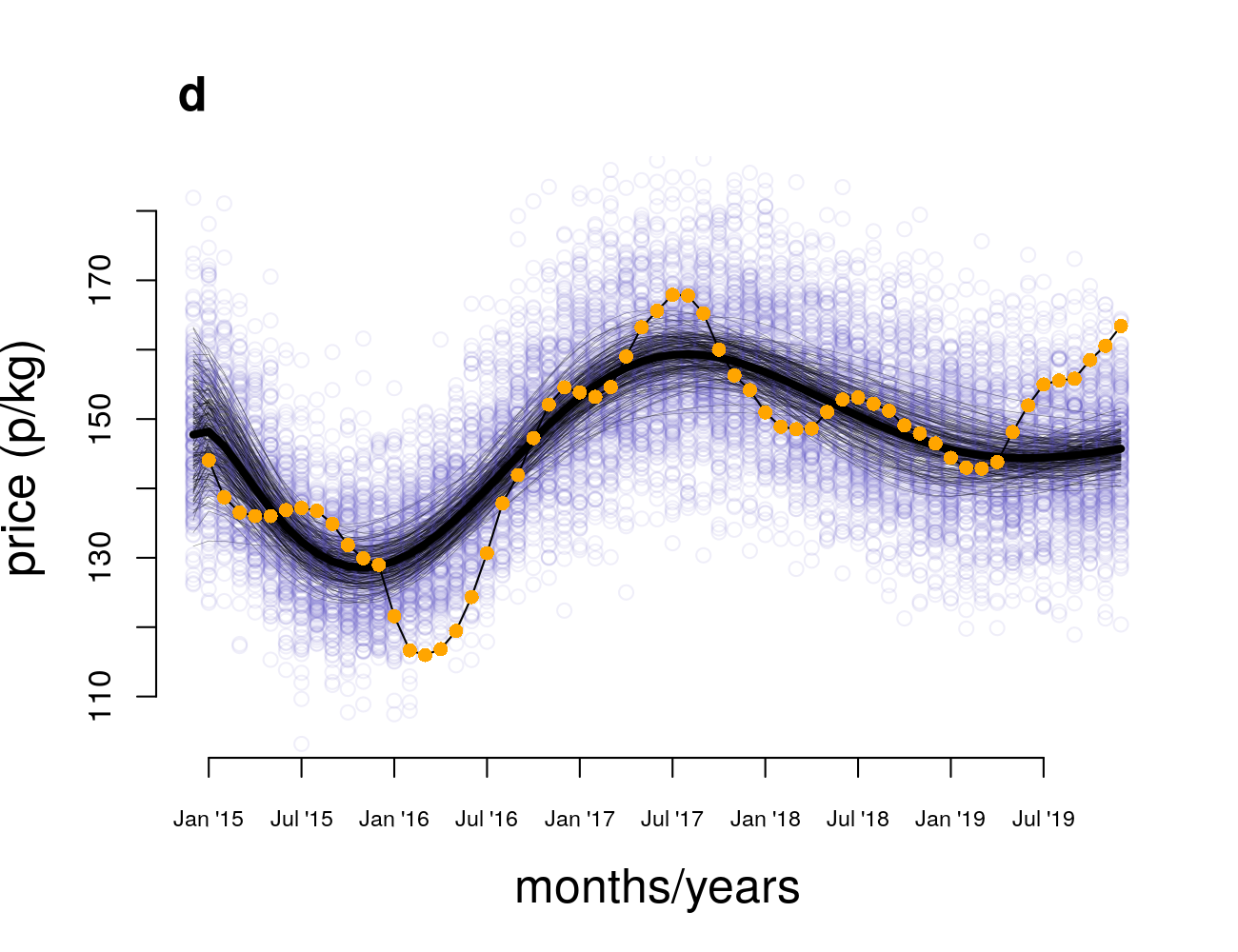}
  \end{subfigure}

  \begin{subfigure}{0.5\textwidth}
    \includegraphics[scale=0.5]{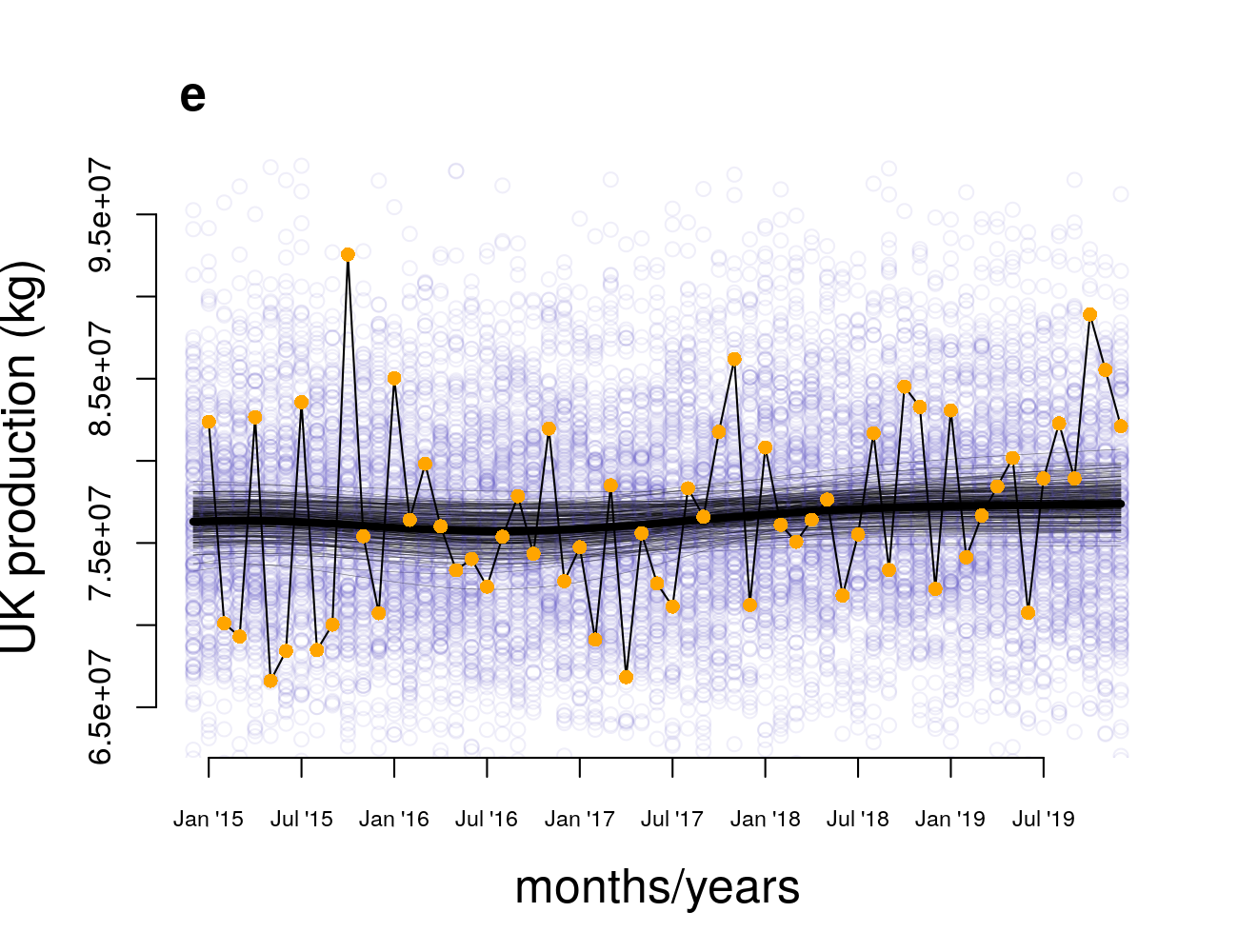}
  \end{subfigure}%
  \begin{subfigure}{0.5\textwidth}
    \includegraphics[scale=0.5]{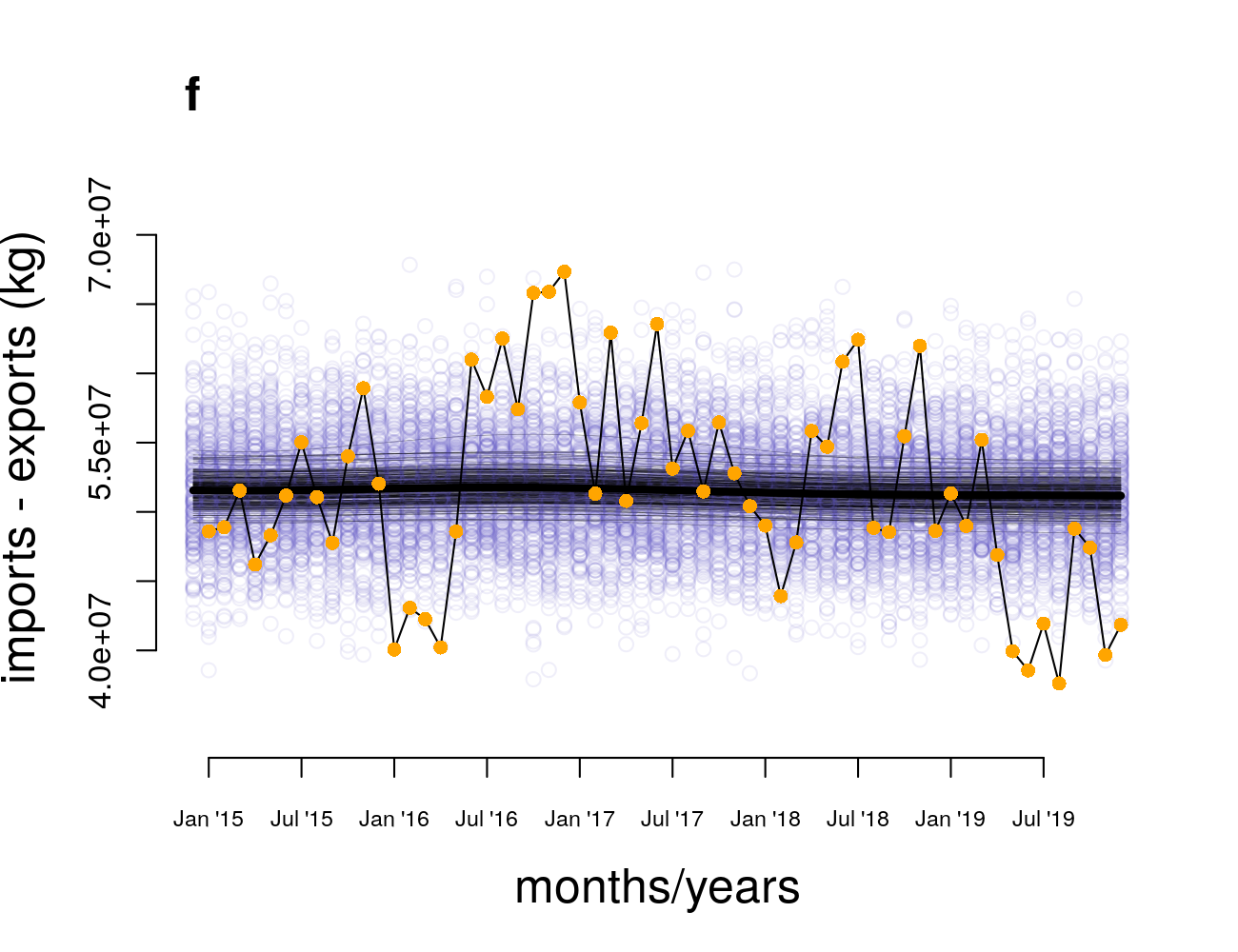}
  \end{subfigure}
  \caption{Fitting the food systems model to the UK pig industry data. Orange filled circles show the raw monthly data (some data is missing), thin black lines display 200 random samples from the posterior distribution, the thick black lines indicate the mean posterior trajectory, and open blue circles display 200 random samples from the posterior predictive distribution (i.e. predictions incorporating random noise ).}
  \label{fig_posterior_predictions}
\end{figure}

Posterior predictions (Figure \ref{fig_posterior_predictions}) reflect the most plausible trajectories of the food system model generating the data, and posterior predictive distributions (open blue circles) cover a large proportion of the observed data. However, there are additional sources of variation that the model trajectories do not account for. For instance, there is seasonal variation in UK pork production: the breeding herd tends to be higher in the July than in the December censuses (Figure \ref{fig_posterior_predictions}a), resulting in higher UK production of pig meat (Figure \ref{fig_posterior_predictions}e) in the latter portions of the year, likely in preparation for the Christmas period. The pig price is also more variable than the model can explain (Figure \ref{fig_posterior_predictions}d), showing a notable drop in 2016 (corresponding to a fall in the EU pig price) and an increase in 2019. The difference between imports and exports (Figure \ref{fig_posterior_predictions}f) fluctuated over the 2015-2019 time period, whereas the model only considers a simple international trade function (equation \ref{eq_inventory}).

\begin{table}[t!]
  \centering
  \footnotesize
  \begin{tabular}{p{5cm}p{3cm}p{5cm}p{2cm}}
    \textbf{Parameter} & \textbf{Mean} & \textbf{95\% HDI} & \textbf{ESS} \\ \hline
    $a$ & 0.0086 & [0, 0.0195] & 5,799 \\
    $e$ & 0.0002 & [0, 0.0007] & 8,175 \\
    $f$ & 2.2712 & [2.2152, 2.3276] & 7,366 \\
    $k$ & 0.3602 & [0.3474, 0.3739] & 7,180 \\
    $h$ & 219478906 & [209862509, 229180910] & 7,320 \\
    $w$ & 0.2392 & [0.0634, 0.4037] & 3,431 \\
    $m$ & 0.0937 & [0.0644, 0.1224] & 5,120 \\
    $q$ & 132.0101 & [102.3935, 161.8261] & 3,502 \\
    $r$ & 0.1514 & [0.0905, 0.2221] & 6,829 \\
    $s$ & 0.6703 & [0.5247, 0.8303] & 3,578 \\
    critical ratio & 1.6796 & [1.5549, 1.7797] & 5,349 \\
  \end{tabular}
  \caption{Key parameter estimates (mean and 95\% highest density interval, HDI) and effective sample sizes (ESS) from fitting the model to the UK pig industry data. Parameters $b$ and $g$ were fixed at the constants 138.3 p/kg and 82.4 kg/pig, respectively.}
  \label{table_parameter_estimates}
\end{table}

\section{Discussion}
This paper has presented a theoretical model of a complex food system that balances analytical tractability and realism. The model represents the functioning of a national food system including international trade, and we have shown that the sustainability of the domestic industry depends on a critical compound parameter that comprises the profitability of the domestic industry (the reference price to cost of capital production ratio), the need for new commodity (commodity waste rates and reference coverage), the ability to produce new capital (captial growth and depreciation rates), and the strength of international trade (see Figure \ref{figure2}a). Below unity, this critical ratio signals that international trade outcompetes the domestic industry and the model enters an unsustainable domestic supply regime characterised by complete reliance on imports. This unsustainable regime also results in higher equilibrium commodity prices, lower inventory and lower consumer demand than when domestic supply is sustainable. By estimating the key parameters of this ratio from data on real food systems, the sustainability of domestic industries, conditional on the assumptions of the model, can be evaluated.

Within the sustainable regime of the model, a key factor determining the long-term behaviour of the food system depends on whether it is characterised by net imports or net exports. In the context of the mathematical model, this is represented by whether the nation produces enough food to meet the reference demand ($h$ in Table \ref{t_symbols}). We find that a food system that must supplement domestic supply with imports in order to meet demand (a net importer) is more vulnerable to collapse because increasing trade potential (higher $k$ values) results in reduced self-sufficiency. Food systems that produce a surplus of domestic commodity can benefit from increased trade potential by exporting more (Figure \ref{figure2}), assuming that export markets are always available. This supports the current literature on the importance of diversifying food commodity sources to ensure food system resilience. Rapid globalisation has meant that 23\% of the food produced globally, and 26\% of global calorie production, is traded \cite{dodorico2014,tu2019,poppy2019}, and the majority of the world's diet is partly dependent on food imports \cite{kummu2020} with several countries not producing enough food to satisfy basic, per-capita caloric intake \cite{dodorico2014}. While global food trade has led to, and encourages further, diversification of diets, it has also led to a less resilient global food system. This is because most countries in the trade network rely on imports from a smaller number of dominant, trade partners \cite{kummu2020}, results supported also by theoretical modelling \cite{tu2019}. The model here provides a critical boundary where a nation's food system may become entirely dependent on imports, leading to higher commodity prices, lower consumer demand and lower overall inventory levels. By the same token, complete self-sufficiency is an unrealistic and potentially harmful goal where the food system is again only reliant on a single supply chain \cite{helm2017}.

Application of our model to 2015-2019 data from the UK pork industry demonstrated that the industry is in the sustainable model regime, with the critical ratio estimated credibly above 1 (Table \ref{table_parameter_estimates}). While the UK pork industry has diminished in size over the last 20 years (by around 50\% since the late 1990s) its current level of self-sufficiency is around 60-65\% and its export market continues to grow due to lower production levels of the Chinese pork industry (e.g. \cite{AHDBexports2020}). The results from our model support this state of the industry: the `trade strength' parameter ($\kappa$), which represents the proportion of the difference between reference demand levels and domestic production that can be traded, was estimated between 35 and 37\%. Crucially, the critical value of $\kappa$ that would tip the UK industry into collaspe is estimated to be 61\%, with a 95\% highest density interval between 56 and 65\%, suggesting if self-sufficiency drops below 50\%, the industry will be closely approaching unsustainability. The UK food system faces a number of challenges in the coming years, including the impact of no-deal Brexit on trade tariffs allowing retailers easier access to cheaper imported meat \cite{feng2017}, the continuing COVID-19 pandemic on the production and processing efficiency of food commodities \cite{power2020}, increasing popularity of reduced meat, vegetarian, vegan and plant-based diets on UK meat demand \cite{james2020}, and the potential for an African Swine Fever epidemic in the pig herd \cite{normile2019,mason2020}. While under some scenarios, such as a bespoke Brexit trade deal with the EU, the price of pig meat might increase \cite{feng2017}, representing a boost to domestic producers, our model indicates that this result is also consistent with a food system coming closer to the critical boundary delineating collapse (i.e. there is a negative relationship between long term price and the critical ratio) and, thus, might not be an advantage to the food system as a whole.

The mechanisms encoded in the model presented here are simple relative to the multi-factorial functioning of real food systems \cite{ericksen2008,ingram2011}. For instance, we have assumed that commodity prices respond only to changes in the supply-demand balance and not extenal factors such as global commodity prices and costs of production. Moreover, the model assumes that international trade responds only to the difference between reference demand and domestic production, and that exports of domestic produce are always available, ignoring how government regulations of trade flows may disrupt this scenario (e.g. pigmeat from pigs fed with feed additives such as ractopomine might fail government import regulations). Our specification that domestic capital changes proportional to the commodity price--production cost ratio \cite{sterman2000} ignores heterogeneity in the structure of food supply chains. While similar assumptions have been used to build more complex system dynamics models (e.g. \cite{meadows1971,sterman2000}), there is great scope for expanding our system of differential equations to reveal the impact of, for instance, dis-aggregated actors of the supply chain (e.g. breeding pigs versus slaughter pigs, producers versus processors), the causal effects of non-finanical drivers (e.g. preserving pig health and welfare) on supply chain functioning, heterogeneity in the production, demand and trade of different product types (e.g. fresh pork versus bacon and sausages), or how external factors, such as the climate, will influence production and demand \cite{vermeulen2012}. Nonetheless, our model provides a description of a single food system supported by both past theoretical and empirical work, and thus offers key insights into the conditions that promote sustainability of domestic food industries.

\subsubsection*{Acknowledgments}
This research is a part of the PigSustain project which is funded through the Global Food Security's `Resilience of the UK Food System Programme', with support from BBSRC, ESRC, NERC and Scottish Government (grant number BB/N020790/1). We thank the Department for Environment, Food and Rural Affairs, and the Agriculture and Horticulture Development Board for making the data used in this article freely accessible.

\subsubsection*{Author contributions}
CG conceptualised the paper, developed and analysed the model, wrote the computer code, analysed the data, and wrote the first draft of the manuscript; SP reviewed model development and contributed to writing and reviewing the manuscript; WHMJ reviewed and contributed to writing the manuscript; NL reviewed and contributed to writing the manuscript; FS reviewed and contributed to writing the manuscript; LMC attained funding, helped conceptualise the model and paper, and reviewed and contributed to writing the manuscript.

\subsubsection*{Data and code accessibility}
All data and code to reproduce the results of this article are available at \href{https://github.com/cmgoold/cfs-model}{https://github.com/cmgoold/cfs-model}.

\bibliographystyle{abbrv}
\bibliography{Goold_et_al_2020-preprint}

\begin{thebibliography}{10}

\bibitem{AHDBeuroexhange2015}
{Agricultural and Horticultural Development Board}.
\newblock {How does the exchange rate affect the GB pig price?}, 2015.
\newblock Available at
  \url{https://pork.ahdb.org.uk/prices-stats/news/2015/august/how-does-the-exchange-rate-affect-the-gb-pig-price/}.

\bibitem{AHDBpocketbook2018}
{Agricultural and Horticultural Development Board}.
\newblock {Pig Pocketbook 2018}, 2018.
\newblock Available at
  \url{https://ahdb.org.uk/knowledge-library/pig-and-poultry-pocketbook}.

\bibitem{AHDBpigmeatrade}
{Agricultural and Horticultural Development Board}.
\newblock {Pig meat trade}, 2020.
\newblock Available at \url{https://ahdb.org.uk/pork/pig-meat-trade}.

\bibitem{AHDBexports2020}
{Agricultural and Horticultural Development Board}.
\newblock {UK pig meat exports show strong growth in July}, 2020.
\newblock Available at
  \url{https://ahdb.org.uk/news/uk-pig-meat-exports-show-strong-growth-in-july}.

\bibitem{allen2016}
T.~Allen and P.~Prosperi.
\newblock Modeling sustainable food systems.
\newblock {\em Environmental Management}, 57(5):956--975, 2016.

\bibitem{barrett2010}
C.~B. Barrett.
\newblock Measuring food insecurity.
\newblock {\em Science}, 327(5967):825--828, 2010.

\bibitem{bene2016}
C.~B{\'e}n{\'e}, D.~Headey, L.~Haddad, and K.~von Grebmer.
\newblock Is resilience a useful concept in the context of food security and
  nutrition programmes? {S}ome conceptual and practical considerations.
\newblock {\em Food Security}, 8(1):123--138, 2016.

\bibitem{bowman2013}
A.~Bowman, J.~Froud, S.~Johal, A.~Leaver, and K.~Williams.
\newblock Opportunist dealing in the uk pig meat supply chain: trader
  mentalities and alternatives.
\newblock {\em Accounting Forum}, 37(4):300--314, 2013.

\bibitem{boyd2003}
R.~Boyd, H.~Gintis, S.~Bowles, and P.~J. Richerson.
\newblock The evolution of altruistic punishment.
\newblock {\em Proceedings of the National Academy of Sciences},
  100(6):3531--3535, 2003.

\bibitem{carpenter2017}
B.~Carpenter, A.~Gelman, M.~D. Hoffman, D.~Lee, B.~Goodrich, M.~Betancourt,
  M.~Brubaker, J.~Guo, P.~Li, and A.~Riddell.
\newblock Stan: a probabilistic programming language.
\newblock {\em Journal of Statistical Software}, 76(1), 2017.

\bibitem{chung2018}
B.~Chung.
\newblock System dynamics modelling and simulation of the {M}alaysian rice
  value chain: Effects of the removal of price controls and an import monopoly
  on rice prices and self-sufficiency levels in {M}alaysia.
\newblock {\em Systems Research and Behavioral Science}, 35(3):248--264, 2018.

\bibitem{coase1935}
R.~H. Coase and R.~F. Fowler.
\newblock Bacon production and the pig-cycle in {G}reat {B}ritain.
\newblock {\em Economica}, 2(6):142--167, 1935.

\bibitem{dawson2009}
P.~J. Dawson.
\newblock The {UK} pig cycle: a spectral analysis.
\newblock {\em British Food Journal}, 111(11), 2009.

\bibitem{deGoede2013}
D.~De~Goede, B.~Gremmen, and M.~Blom-Zandstra.
\newblock Robust agriculture: balancing between vulnerability and stability.
\newblock {\em NJAS-Wageningen Journal of Life Sciences}, 64:1--7, 2013.

\bibitem{DEFRA2019auk18}
{Department for Environment, Food and Rural Affairs}.
\newblock {Agriculture in the United Kingdom 2018}, 2019.

\bibitem{DEFRApigcattlestats2020}
{Department for Environment, Food and Rural Affairs}.
\newblock {Latest cattle, sheep and pig slaughter statistics}, 2020.
\newblock Available at
  \url{https://www.gov.uk/government/statistics/cattle-sheep-and-pig-slaughter}.

\bibitem{DEFRAlivestocknumbers}
{Department for Environment, Food and Rural Affairs}.
\newblock {Livestock numbers in England and the UK}, 2020.
\newblock Available at
  \url{https://www.gov.uk/government/statistical-data-sets/structure-of-the-livestock-industry-in-england-at-december}.

\bibitem{DEFRAlivestockprices}
{Department for Environment, Food and Rural Affairs}.
\newblock {Livestock prices, finished and store}, 2020.
\newblock Available at
  \url{https://www.gov.uk/government/statistical-data-sets/livestock-prices-finished-and-store}.

\bibitem{dodorico2014}
P.~D'Odorico, J.~A. Carr, F.~Laio, L.~Ridolfi, and S.~Vandoni.
\newblock Feeding humanity through global food trade.
\newblock {\em Earth's Future}, 2(9):458--469, 2014.

\bibitem{drimie2013}
S.~Drimie and M.~McLachlan.
\newblock Food security in {S}outh {A}frica: first steps toward a
  transdisciplinary approach.
\newblock {\em Food Security}, 5(2):217--226, 2013.

\bibitem{ericksen2010}
P.~Ericksen, B.~Stewart, J.~Dixon, D.~Barling, P.~Loring, M.~Anderson, and
  J.~Ingram.
\newblock The value of a food system approach.
\newblock {\em Food security and global environmental change}, 25:24--25, 2010.

\bibitem{ericksen2008}
P.~J. Ericksen.
\newblock Conceptualizing food systems for global environmental change
  research.
\newblock {\em Global environmental change}, 18(1):234--245, 2008.

\bibitem{BPEXprofitability2011}
B.~P. Executive.
\newblock {Profitability in the pig supply chain}.
\newblock Technical report, 2011.

\bibitem{ezekiel1938}
M.~Ezekiel.
\newblock The cobweb theorem.
\newblock {\em The Quarterly Journal of Economics}, 52(2):255--280, 1938.

\bibitem{feng2017}
S.~Feng, M.~Patton, J.~Binfield, and J.~Davis.
\newblock `{D}eal' or `no deal'? {I}mpacts of alternative post-brexit trade
  agreements on uk agriculture.
\newblock {\em EuroChoices}, 16(3):27--33, 2017.

\bibitem{FAO2009}
{Food and Agriculture Organization}.
\newblock {\em World summit on food security: declaration of the world summit
  on food security}.
\newblock FAO, 2009.
\newblock Available at \url{http://www.fao.org/wsfs/wsfs-list-documents/en/}.

\bibitem{gao2016}
J.~Gao, B.~Barzel, and A.-L. Barab{\'a}si.
\newblock Universal resilience patterns in complex networks.
\newblock {\em Nature}, 530(7590):307, 2016.

\bibitem{gouel2012}
C.~Gouel.
\newblock Agricultural price instability: a survey of competing explanations
  and remedies.
\newblock {\em Journal of economic surveys}, 26(1):129--156, 2012.

\bibitem{haldane1934}
J.~Haldane.
\newblock A contribution to the theory of price fluctuations.
\newblock {\em The Review of Economic Studies}, 1(3):186--195, 1934.

\bibitem{hammond2012}
R.~A. Hammond and L.~Dub{\'e}.
\newblock A systems science perspective and transdisciplinary models for food
  and nutrition security.
\newblock {\em Proceedings of the National Academy of Sciences},
  109(31):12356--12363, 2012.

\bibitem{harlow1960}
A.~A. Harlow.
\newblock The hog cycle and the cobweb theorem.
\newblock {\em Journal of Farm Economics}, 42(4):842--853, 1960.

\bibitem{helm2017}
D.~Helm.
\newblock Agriculture after {B}rexit.
\newblock {\em Oxford Review of Economic Policy}, 33(suppl\_1):S124--S133,
  2017.

\bibitem{ingram2011}
J.~S. Ingram.
\newblock A food systems approach to researching food security and its
  interactions with global environmental change.
\newblock {\em Food Security}, 3(4):417--431, 2011.

\bibitem{james2020}
W.~H. James, N.~Lomax, M.~Birkin, and L.~M. Collins.
\newblock Geodemographic patterns of meat expenditure in {G}reat {B}ritain.
\newblock {\em Applied Spatial Analysis and Policy}, x:1--28, 2020.

\bibitem{kermack1927}
W.~O. Kermack and A.~G. McKendrick.
\newblock A contribution to the mathematical theory of epidemics.
\newblock {\em Proceedings of the Royal Society of London A},
  115(772):700--721, 1927.

\bibitem{kummu2020}
M.~Kummu, P.~Kinnunen, E.~Lehikoinen, M.~Porkka, C.~Queiroz, E.~R{\"o}{\"o}s,
  M.~Troell, and C.~Weil.
\newblock Interplay of trade and food system resilience: gains on supply
  diversity over time at the cost of trade independency.
\newblock {\em Global Food Security}, 24:100360, 2020.

\bibitem{legrand2019}
N.~Legrand.
\newblock The empirical merit of structural explanations of commodity price
  volatility: review and perspectives.
\newblock {\em Journal of Economic Surveys}, 33(2):639--664, 2019.

\bibitem{marchand2016}
P.~Marchand, J.~A. Carr, J.~Dell’Angelo, M.~Fader, J.~A. Gephart, M.~Kummu,
  N.~R. Magliocca, M.~Porkka, M.~J. Puma, Z.~Ratajczak, et~al.
\newblock Reserves and trade jointly determine exposure to food supply shocks.
\newblock {\em Environmental Research Letters}, 11(9):095009, 2016.

\bibitem{mason2020}
D.~Mason-D'Croz, J.~R. Bogard, M.~Herrero, S.~Robinson, T.~B. Sulser, K.~Wiebe,
  D.~Willenbockel, and H.~C.~J. Godfray.
\newblock Modelling the global economic consequences of a major {A}frican swine
  fever outbreak in {C}hina.
\newblock {\em Nature Food}, 1(4):221--228, 2020.

\bibitem{maxwell1996}
S.~Maxwell.
\newblock Food security: a post-modern perspective.
\newblock {\em Food policy}, 21(2):155--170, 1996.

\bibitem{may1973}
R.~M. May.
\newblock Qualitative stability in model ecosystems.
\newblock {\em Ecology}, 54(3):638--641, 1973.

\bibitem{meadows1971}
D.~L. Meadows.
\newblock {\em Dynamics of commodity production cycles}.
\newblock PhD thesis, Massachusetts Institute of Technology, Massachusetts
  Institute of Technology, 1969.

\bibitem{nerlove1958}
M.~Nerlove.
\newblock Adaptive expectations and cobweb phenomena.
\newblock {\em The Quarterly Journal of Economics}, 72(2):227--240, 1958.

\bibitem{ngonghala2017}
C.~N. Ngonghala, G.~A. De~Leo, M.~M. Pascual, D.~C. Keenan, A.~P. Dobson, and
  M.~H. Bonds.
\newblock General ecological models for human subsistence, health and poverty.
\newblock {\em Nature ecology \& evolution}, 1(8):1153--1159, 2017.

\bibitem{normile2019}
D.~Normile.
\newblock African swine fever marches across much of {A}sia, 2019.

\bibitem{nystrom2019}
M.~Nystr{\"o}m, J.-B. Jouffray, A.~V. Norstr{\"o}m, B.~Crona, P.~S.
  J{\o}rgensen, S.~Carpenter, {\"O}.~Bodin, V.~Galaz, and C.~Folke.
\newblock Anatomy and resilience of the global production ecosystem.
\newblock {\em Nature}, 575(7781):98--108, 2019.

\bibitem{ottoday2011}
S.~P. Otto and T.~Day.
\newblock {\em A biologist's guide to mathematical modeling in ecology and
  evolution}.
\newblock Princeton University Press, 2011.

\bibitem{otto2020}
S.~P. Otto and A.~Rosales.
\newblock Theory in service of narratives in evolution and ecology.
\newblock {\em The American Naturalist}, 195(2):290--299, 2020.

\bibitem{parker2014}
P.~S. Parker and J.~Shonkwiler.
\newblock On the centenary of the {German} hog cycle: new findings.
\newblock {\em European Review of Agricultural Economics}, 41(1):47--61, 2014.

\bibitem{pimm1984}
S.~L. Pimm.
\newblock The complexity and stability of ecosystems.
\newblock {\em Nature}, 307(5949):321--326, 1984.

\bibitem{poppy2019}
G.~M. Poppy, J.~Baverstock, and J.~Baverstock-Poppy.
\newblock Meeting the demand for meat: analysing meat flows to and from the
  {UK} pre and post brexit.
\newblock {\em Trends in Food Science \& Technology}, 86:569--578, 2019.

\bibitem{power2020}
M.~Power, B.~Doherty, K.~Pybus, and K.~Pickett.
\newblock How covid-19 has exposed inequalities in the uk food system: The case
  of uk food and poverty.
\newblock {\em Emerald Open Research}, 2, 2020.

\bibitem{rcoreteam2020}
{R Core Team}.
\newblock {\em R: A Language and Environment for Statistical Computing}.
\newblock R Foundation for Statistical Computing, Vienna, Austria, 2020.

\bibitem{sampedro2020}
C.~Sampedro, F.~Pizzitutti, D.~Quiroga, S.~J. Walsh, and C.~F. Mena.
\newblock Food supply system dynamics in the {G}alapagos islands: agriculture,
  livestock and imports.
\newblock {\em Renewable Agriculture and Food Systems}, 35(3):234--248, 2020.

\bibitem{scalco2019}
A.~Scalco, J.~I. Macdiarmid, T.~Craig, S.~Whybrow, and G.~W. Horgan.
\newblock An agent-based model to simulate meat consumption behaviour of
  consumers in {B}ritain.
\newblock {\em Journal of Artificial Societies and Social Simulation}, 22(4):8,
  2019.

\bibitem{scheffer2001}
M.~Scheffer, S.~Carpenter, J.~A. Foley, C.~Folke, and B.~Walker.
\newblock Catastrophic shifts in ecosystems.
\newblock {\em Nature}, 413(6856):591--596, 2001.

\bibitem{seekell2017}
D.~Seekell, J.~Carr, J.~Dell’Angelo, P.~D’Odorico, M.~Fader, J.~Gephart,
  M.~Kummu, N.~Magliocca, M.~Porkka, M.~Puma, et~al.
\newblock Resilience in the global food system.
\newblock {\em Environmental Research Letters}, 12(2):025010, 2017.

\bibitem{smaldino2019}
P.~Smaldino.
\newblock Better methods can't make up for mediocre theory.
\newblock {\em Nature}, 575(7781):9, 2019.

\bibitem{smaldino2017}
P.~E. Smaldino.
\newblock Models are stupid, and we need more of them.
\newblock In R.~Vallacher, S.~Read, and A.~Nowak, editors, {\em Computational
  Models in Social Psychology}, pages 311--331. Routledge New York, NY, 2017.

\bibitem{sole1996}
R.~V. Sol{\'e}, S.~Manrubia~Cuevas, B.~Luque, J.~Delgado, and J.~Bascompte.
\newblock Phase transitions and complex systems: simple, nonlinear models
  capture complex systems at the edge of chaos.
\newblock {\em Complexity}, 1, 1996.

\bibitem{springmann2018}
M.~Springmann, M.~Clark, D.~Mason-D’Croz, K.~Wiebe, B.~L. Bodirsky,
  L.~Lassaletta, W.~de~Vries, S.~J. Vermeulen, M.~Herrero, K.~M. Carlson,
  et~al.
\newblock Options for keeping the food system within environmental limits.
\newblock {\em Nature}, 562(7728):519, 2018.

\bibitem{stan2019}
{Stan Development Team}.
\newblock {RStan}: the {R} interface to {Stan}, 2019.
\newblock R package version 2.19.2.

\bibitem{sterman2000}
J.~Sterman.
\newblock {\em Business dynamics}.
\newblock Irwin/McGraw-Hill, 2000.

\bibitem{strogatz1994}
S.~H. Strogatz.
\newblock {\em Nonlinear Dynamics and Chaos}.
\newblock CRC Press, 1994.

\bibitem{strzepek2010}
K.~Strzepek and B.~Boehlert.
\newblock Competition for water for the food system.
\newblock {\em Philosophical Transactions of the Royal Society B: Biological
  Sciences}, 365(1554):2927--2940, 2010.

\bibitem{suweis2015}
S.~Suweis, J.~A. Carr, A.~Maritan, A.~Rinaldo, and P.~D'Odorico.
\newblock Resilience and reactivity of global food security.
\newblock {\em Proceedings of the National Academy of Sciences},
  112(22):6902--6907, 2015.

\bibitem{taylor2006}
D.~H. Taylor.
\newblock Strategic considerations in the development of lean agri-food supply
  chains: a case study of the {UK} pork sector.
\newblock {\em Supply Chain Management: An International Journal}, 11(3), 2006.

\bibitem{tendall2015}
D.~Tendall, J.~Joerin, B.~Kopainsky, P.~Edwards, A.~Shreck, Q.~Le,
  P.~Kr{\"u}tli, M.~Grant, and J.~Six.
\newblock Food system resilience: defining the concept.
\newblock {\em Global Food Security}, 6:17--23, 2015.

\bibitem{tu2019}
C.~Tu, S.~Suweis, and P.~D'Odorico.
\newblock Impact of globalization on the resilience and sustainability of
  natural resources.
\newblock {\em Nature Sustainability}, 2(4):283, 2019.

\bibitem{vermeulen2012}
S.~J. Vermeulen, B.~M. Campbell, and J.~S. Ingram.
\newblock Climate change and food systems.
\newblock {\em Annual review of environment and resources}, 37, 2012.

\bibitem{zawadzka2010}
D.~Zawadzka.
\newblock The history of research on the `pig cycle'.
\newblock Problems of Agricultural Economics / Zagadnienia Ekonomiki Rolnej
  205144, Institute of Agricultural and Food Economics - National Research
  Institute (IAFE-NRI), 2010.

\end{thebibliography}
\end{document}